\documentclass[fleqn,usenatbib,useAMS]{mnras}

\usepackage{graphicx}
\usepackage{amsmath}
\usepackage{amssymb}
\usepackage{subfig}
\usepackage{rotating}
\usepackage{times}

\title[Supernova feedback in molecular clouds]{Supernova feedback and the energy deposition in molecular clouds}
\author[W.~E.~Lucas et al.]{William E.~Lucas$^{1,2}$,
 Ian A.~Bonnell$^1$\thanks{E-mail: iab1@st-andrews.ac.uk} \& James E.~Dale$^3$ \\
  $^1$ SUPA, School of Physics \& Astronomy, University of
    St Andrews, North Haugh, St Andrews, Fife KY16 9SS, United Kingdom \\
  $^2$ EPCC, Bayes Centre, 47 Potterrow, Edinburgh, EH8 9BT, United Kingdom \\
  $^3$ School of Physics Astronomy and Mathematics, University of Hertfordshire, Hatfield, Hertfordshire, United Kingdom}
  
\begin{document}

\maketitle

\begin{abstract}  
Feedback from supernovae is often invoked as an important process in limiting star formation, removing gas from galaxies and hence as a determining process in galaxy formation. Here we report on numerical simulations investigating the interaction between supernova explosions and the natal molecular cloud. We also consider the cases with and without previous feedback from the high-mass star in the form of ionising radiation and stellar winds.  The supernova is able to find weak points in the cloud and create channels through which it can escape, leaving much of the well shielded cloud largely unaffected. This effect is increased when the channels are pre-existing due to the effects of previous stellar feedback.  The expanding supernova deposits its energy in the gas that is in these exposed channels, and hence sweeps up less mass when feedback has already occurred, resulting in faster outflows with less radiative losses. The full impact of the supernova explosion is then able to impact the larger scale of the galaxy in which it abides. We conclude that supernova explosions only have moderate effects on their dense natal environments but that with pre-existing feedback, the energetic effects of the supernova are able to escape and affect the wider scale medium of the galaxy.  
\end{abstract}

\section{Introduction} \label{s:intro}

Star formation and the feedback from young stars are the primary internal processes that can drive galaxy evolution \citep{SchEA15}. These processes are  interconnected 
as feedback can affect the local star forming environment, as well as the large-scale interstellar medium from which the next generation of stars form \citep{DobbsPPVI}. The morphology and rate of star formation (and therefore stellar feedback) also depend on the large-scale structure and dynamics of the host galaxies. 

Numerical simulations have been instrumental in helping us understand how star formation proceeds and the dynamical processes that contribute to the observed stellar properties from the origin of stellar clusters \citep{BBV2003,SB2017}, the initial mass function \citep{BCBP2001,BLZ2007}, the formation of high-mass stars \citep{BVB2004,Krum2009,ZY2007,SLB2009,KKBH2010} and the origin of  binary and multiple systems \citep{BBB2003,Bate2012,ReipPPVI}.  At the same time, numerical simulations have been used to explore how the initial conditions for star formation arise due to large-scale flows, spiral shocks or other dynamical events in the galaxy \citep{Dobbs2011,BDS2013,DobbsPPVI,SB2016,VSGSC2017}.  

Feedback from young stars, in the form of ionising radiation, stellar winds and ultimately supernova explosions, are often invoked as processes to limit the rate at which stars form, 
recycle molecular gas back into the interstellar medium (ISM) , as well as a significant source of kinetic energy into molecular clouds and the ISM \citep{KrumPPVI}. On larger scales, supernova explosions are considered essential in regulating star formation and gas expulsion in galaxies,  helping to explain the  stellar mass function of galaxies \citep{SchEA15}.

Including ionisation and stellar wind feedback on smaller scales has shown that the local gas environment plays a crucial role in channeling the outflows through weakpoints in the surrounding clouds \citep{DBCB2005,DEB2012}, allowing the feedback energy to largely escape and leaving more massive clouds less affected than would be estimated \citep{DEA14,Dale2017,Ali2018}. Similar conclusions arise when supernova feedback is invoked in dense, structured environments \citep{RogersPittard2013,Kortgen2016} although in many studies  resolution issues necessitate lower density, near-uniform environments and/or subgrid  treatments  \citep{Agertz2013,Walch2015,Geen2015,ReyRaposo2017}.

Including supernova feedback on galactic scales \citep{YEA97,MW03,SSS18} has generally necessitated relying on subgrid physics or on under-resolved simulations. Issues that arise include uncertainty over tuning the subgrid physics \citep{REA17}, and initial losses (e.g. overcooling) due to the lack of resolution 
that artificially reduce the uptake of supernova energy by the ISM \citep*{KWH96,DVS12}.

In this paper, we present work investigating the effects of supernova explosions in a realistic star-forming molecular cloud, and how the supernova's evolution is affected by previous stellar feedback in the form of ionising radiation and stellar winds. 

\section{Numerical methods}

\subsection{Smoothed particle hydrodynamics simulations}

To perform our simulations we used the smoothed particle hydrodynamics (SPH) code \textsc{sphNG} (\citealt{BE90}; \citealt*{BBP95}). Smoothed particle hydrodynamics is a Lagrangian method for computational fluid dynamics. The fluid is represented as a set of particles whose properties (positions, velocities, internal energies, densities, etc) evolve according to fluid equations calculated over neighbouring particles using a smoothing kernel.  Each particle is itself smoothed through the surrounding volume defined by its kernel, allowing interpolated fluid quantities to be determined at any point in space. \citet{BE90} and \citet{MO92} provide useful introductory overviews of SPH. 

Particles in the simulation were evolved on individual timesteps \citep{HK89}, but this is known to be a source of error in energy and momentum conservation \citep{SM09}. The problem is most pronounced when particles on very different timesteps interact with one another: the timestepping scheme may not allow particles with long timesteps, such as those representing slowly evolving ambient gas, to ever `feel' the effects of passing short timestep particles, such as those in SN ejecta, which are themselves receiving energy and momentum contributions. To circumvent this problem, a development of the timestep limiter scheme similar to that described by \citet{SM09} and \citet{DDV12} was employed. When deployed within the simulation code's Runge-Kutta-Fehlberg integrator, this maintained a factor of at most $4$ between neighbour particles' timesteps to ensure momentum and energy conservation. The timestep limiter also includes an immediate reduction of timesteps when neighbouring particles violate the above condition, as  can occur in strong shocks.

We used the Run I simulations of \citet{DEA14} as the basis for our supernova simulations. To briefly describe Run I, it used $10^6$ particles to simulate a spherical centrally concentrated cloud of $10^4 M_\odot$. The SPH gas particle mass was thus $0.01 M_\odot$. The cloud was $10\,\mathrm{pc}$ in radius and turbulently supported with an initial ratio of energies ${E_{\rm kin}} / {|E_{\rm grav}|} = 0.7$. The cloud was then allowed to evolve using the ionization scheme of \citet*{DEC07} and the winds of \citet{DB08}. Dynamically created sink particles \citep{BBP95} were used to represent forming stars.

Three variations on Run I were used as the initial points for our supernova simulations. One, referred to henceforth as `DFB' for dual feedback, had been run to the insertion point using the full feedback model including both ionization and winds from the massive stars which formed. Another, `ION', used only ionizing feedback. `NFB' was a control run which had included neither form of feedback and thus was evolved with hydrodynamics under gravity alone. For each of these  initial conditions, two versions were run, one including the supernova (with run names postfixed `-S') and another control run without (postfixed `-N'). This allowed us to isolate the effects of the supernova in each run by also following the events that would take place in its absence.

\subsection{Supernova method} \label{ss:sn_method}

In each simulation we inserted the supernova at the position of the most massive sink particle. The ejecta mass was set to $\approx 23.9 M_\odot$, $25 \%$ that of the most massive sink in the original control run. While the most massive sinks in the two runs with feedback were at lower values ($\approx 2/3$ that in the no-feedback run), we opted to use this mass in all three setups in order for the supernovae in each to more closely resemble one another. Using the higher mass progenitor also ensured the highest number of particles in the supernova.

The supernovae ejecta were directly inserted by creating new gas particles around the most massive sink in each simulation. The mass of that sink was decreased by the same amount. The particles were randomly positioned within a sphere of radius $r_\mathrm{SN} = 0.1 \, \mathrm{pc}$, and with a central hole of equal radius to the sink particle accretion radius ($10^{-3}\,\mathrm{pc}$). Any gas particles already inside the SN insertion region were removed and then reinserted alongside the SN particles to conserve mass. A total supernova energy of $10^{51}\,\mathrm{ergs}$ was split between kinetic and thermal forms, with the velocities directed outwards radially from the supernova centre. 

We created supernovae using several recipes for the distribution of energy in the ejecta. We found that using a $90:10$ ratio of kinetic to thermal energy resulted in an exceptionally well-formed shock forming from nearby swept-up material. However, the supernova ejecta itself almost shattered on impact with this material, forming small cannonball-like clumps of approximately 50 particles -- the same as \textsc{sphNG}'s target number of neighbour particles. In the end we settled on a $50:50$ split of kinetic and thermal energy, giving us a total of $5\times 10^{50}$ ergs in each form.


To ensure a well formed shock, the  kinetic energies assigned to the supernova particles followed a $r^{-1}$ radial profile with an inner core. This led the inner regions to catch up to the outer regions of the supernova ejecta, creating a well-formed shell. The thermal energies were  distributed uniformly. 

In order to assign particles velocities according to the this profile we first had to choose the core radius $r_\mathrm{core}$, and define the fractional core size $f_r = r_\mathrm{core}/r_\mathrm{SN}$. We also defined the ratio of specific kinetic energy at the SN edge to that in the core, $f_e = e_k(r_\mathrm{SN})/e_{k,\mathrm{core}}$. 
For our simulations we found $f_e = f_r = 0.6$ worked well in producing a dense expanding shell.

In its modern form \textsc{sphNG} uses the grad-h formulation of SPH (see e.g. \citealt{PM04}), requiring that gas particle masses be constant. 
In order to assure that the simulations accurately captured the interaction between ejecta and environment, we increased the resolution of the full simulation. Thus  before the supernova's insertion we split each original $0.01 \mathrm{M}_\odot$ gas particle into nine, giving a particle mass of $0.00\bar{1}M_\odot$. This increased the number of particles inserted for the supernovae from $\approx 2400$ to $\approx 21500$ (neglecting any re-inserted particles from the nearby medium).

\subsection{Cooling} \label{ss:cooling}
The internal energy of the gas in our simulations was allowed to evolve following the method presented in \citet{VSEA07}. With this method, the timescale required for each particle to reach its equilibrium internal energy was calculated, taking into account radiative and hydrodynamic heating and cooling, and an implicit integration towards equilibrium was then performed. The cooling curve of \citet{KI02} was used for low temperature ($T<10^4$ K) gas. The curve was extended to $10^9\,\mathrm{K}$,
with a uniform cooling rate from $10^4\,\mathrm{K}$ to $2 \times 10^5\,\mathrm{K}$. At  higher temperatures the cooling rate decreases as described by e.g. \citet{DMC72} and \citet{F03}. The final cooling curve is thus similar in form to the solar abundance, $z=0$, $n = 1$ or $100\,\mathrm{cm}^{-3}$ curves of \citet{DREA13}.

\section{Overview of evolution}
The main goals of this study were to investigate how the feedback from the supernova interacted with its natal cloud and how this interaction depended on any previous feedback in terms of ionising radiation and stellar winds. The simulation without  previous feedback, NFB-S, has the stars deeply embedded within the dense molecular gas. In contrast, both simulations with previous stellar feedback, ION-S where ionisation is included and DFB-S where both ionisation and stellar winds are included,  have the high-mass stars and the central regions of the cluster within an HII region. The rarefied nature of the immediate environment are clearly
significantly different from the highly structured nature of the dense gas in the no-feedback run and thus play important roles in shaping how the supernova interacts with the environment and how the supernova energy is able to escape from the cloud. As mentioned before, these initial conditions were derived from the Run I simulations of \citet{DEA14}. We will first discuss the two cases, with and without previous feedback, separately.

\subsection{Evolution without previous feedback (Run NFB-S)}

\begin{figure*}
	\begin{center}
	\includegraphics[width=\textwidth]{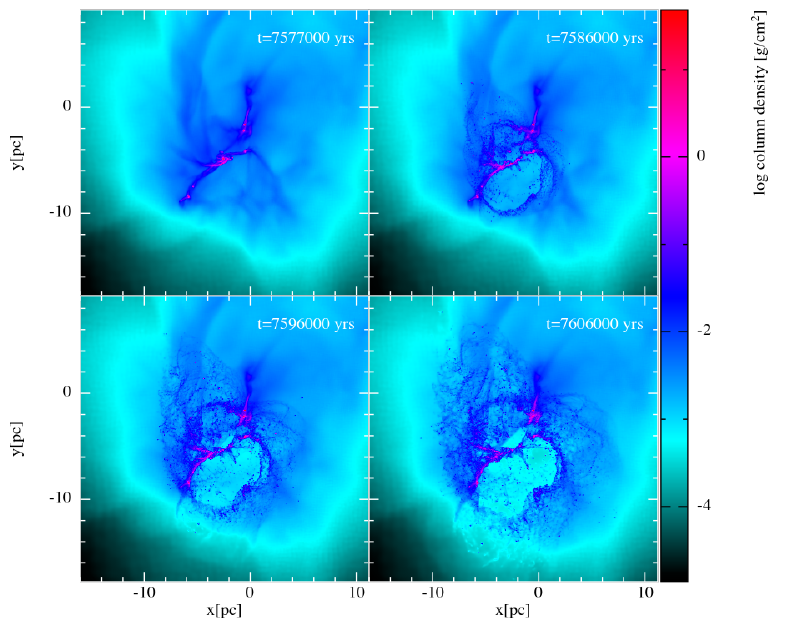}
	\caption{Evolution of run NFB-S, shown as column density in $xy$-projection. The supernova progenitor was located in the densest regions at the centre of the cloud and is visible as the small high density ball in the first panel. Due to its location, the explosion was confined and formed a large expanding bubble. The denser material constrained the explosion more effectively, giving it the appearance of a double-bubble in this projection -- in $xz$- and $yz$-projection this is not the case as the progenitor was offset from the centre. While dense gas survived, its structure was markedly altered after the supernova's passage.}
	\label{fig:nfb_evo}
	\end{center}
\end{figure*}

At the point of supernova detonation in the no-previous-feedback simulation, NFB-S, the progenitor was a member of the star cluster embedded deep within the cloud's dense core. This dense gas remained due to the lack of any kind of previous feedback. It confined the explosion and led to the formation and expansion of a blast wave. Densities of $10^{-21}$ to $10^{-19}\,\mathrm{g}\,\mathrm{cm}^{-1}$ in the vicinity of the explosion mean that the explosion converted from free expansion to the Sedov-Taylor phase within a distance of $1\,\mathrm{pc}$.

The environment was highly inhomogeneous leading to a blast wave that was far from spherically symmetric. In the $xy$-projection, shown in Figure~\ref{fig:nfb_evo}, we see two blasts expanding into lower density gas above and below the central dense region with the appearance of a large filament. This geometry of the feedback  largely resembles the shape of the feedback bubble from ionisation (and ionisation plus winds) in the earlier \citet{DEA14} study. The higher-density filament seen in Figure~\ref{fig:nfb_evo} was significantly more robust to the initial blast wave but affected over longer timescales by the high pressure in the supernova remnant. From other perspectives however, the large filament is revealed to  be the projection of smaller more complex set of filaments within the cloud, and the supernova did indeed expand in a single blast wave, albeit a highly asymmetric one. This supernova then is not too dissimilar from those described by \citet{HEA16} whose model shows a supernova expanding at different rates into cones of different densities.

\begin{figure}
	\begin{center}
		\includegraphics[width=8cm]{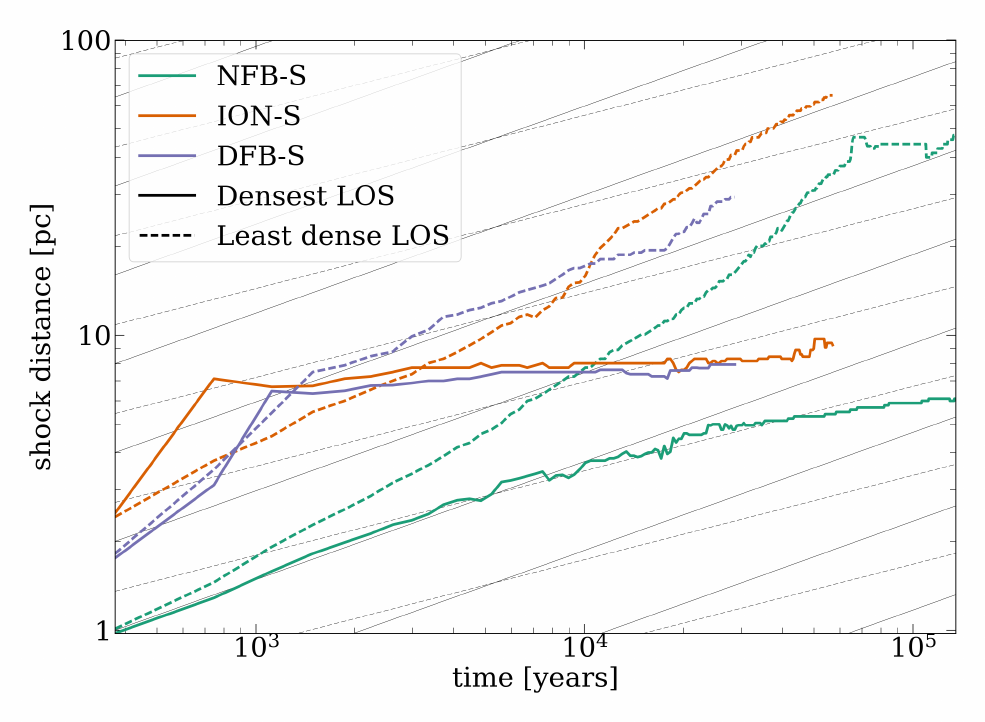} \hspace{0cm}
		\includegraphics[width=8cm]{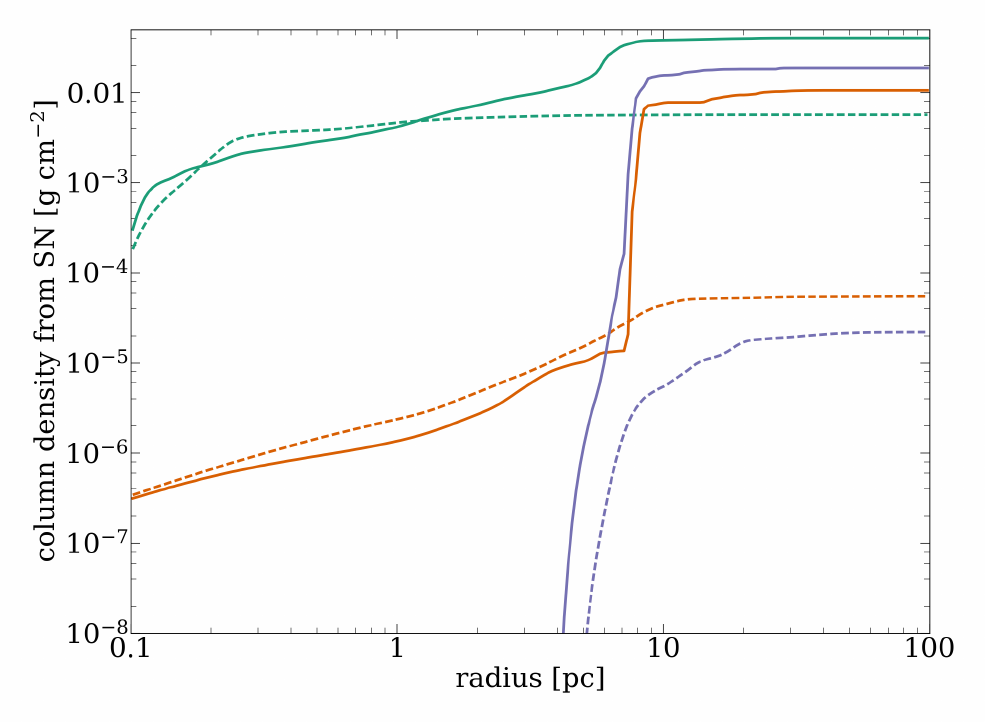}
		\caption{The upper plot shows the expansion of the shock driven by the supernova. The position of the shock is shown along two lines of sight for each simulation -- these were determined by taking the simulations at $t_\mathrm{SN} = 0\,\mathrm{years}$ and finding the most and least dense line of sight (LOS) from all those given by the centres of 768 HEALPix pixels \citep{GEA05}, using the supernova as the origin. The solid and dashed grey lines respectively show the Sedov-Taylor and pressure-driven snowplough expansion rates. The lower plot shows the column density from the SN as a function of distance along each corresponding LOS. The expansion of the SN shock was slowest in NFB-S, and in particular along the densest LOS where it failed to move beyond $6\,\mathrm{pc}$ in over $10^5$ years. At the same time, it moved to almost $50\,\mathrm{pc}$ along the path of least resistance (though it should be noted the method for determining the position of the shock began to break down at around this point). Both runs with previous feedback, ION-S and DFB-S, show that the expansion of the shock was halted along the densest LOS, where, as the second plot shows, the column density jumped at the point of entry to the walls around the central cavity. On the other hand, the SN freely expanded along the paths where the column density grew slowly and there were no sudden jumps in density.}
		\label{fig:rshock}
	\end{center}
\end{figure}

This behaviour can be seen more clearly in Figure~\ref{fig:rshock}. The shock moved forwards slowly along the densest line of sight (LOS) such that it reached a distance of $6\,\mathrm{pc}$ from the progenitor star's position by the end of the simulation. The least dense LOS can be seen in the figure's second panel to have not been substantially less crowded, reaching final column densities only an order of magnitude below those in the densest LOS. Nevertheless the difference was enough that the shock was driven forwards to nearly $50,\mathrm{pc}$. This preferential expansion into the paths of least resistance -- lower density regions -- will become even more prominent when ionisation or stellar winds feedback is included in the pre-supernova evolution.

\subsection{Evolution with previous feedback (Runs ION-S and DFB-S)}


\begin{figure*}
	\begin{center}
	\includegraphics[width=\textwidth]{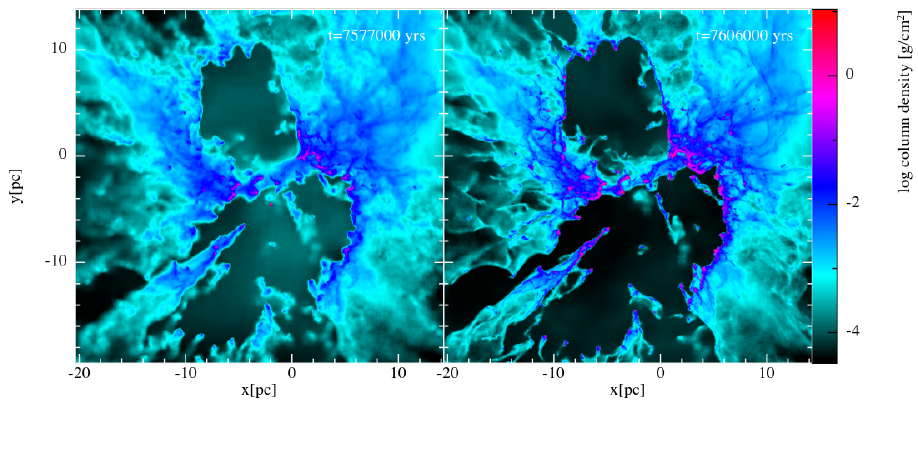}
	\caption{Evolution of run ION-S, showing column density in $xy$-projection at the point of supernova detonation and $2.87 \times 10^4$ years later. The SN is seen as the small high density sphere in the lower cavity (note that this structure is a projection effect only). The changes are not nearly as drastic as those seen in Figure~\ref{fig:nfb_evo}. The SN removed some ionized gas from the cavity, leading to a drop in density, and further sculpted and slightly increased the density in the walls. The cavity itself was slightly larger, most visible in the shape of the topmost cavity walls. The overall effect seems to simply be a continuation of the earlier ionizing feedback's action. Beyond the lower left limits of the plot are an expanding low density shell expelled by the supernova which left the cloud while sweeping up only low density ionized gas.}
	\label{fig:ion_evo}
	\end{center}
\end{figure*}

\begin{figure*}
	\begin{center}
	\includegraphics[width=\textwidth]{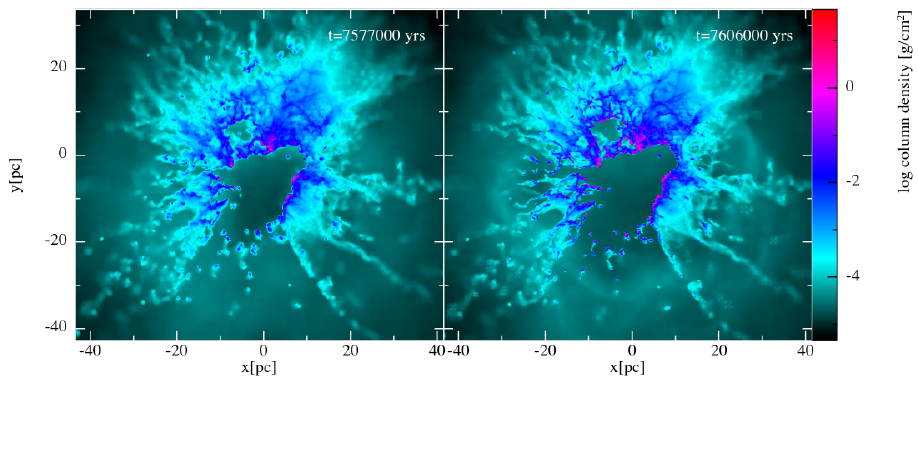}
	\caption{Evolution of run DFB-S, showing column density in $xy$-projection at the point of supernova detonation and $2.87\times  10^4$ years later. The SN is just visible as the small high density sphere in the central cavity. This view is more zoomed out than that shown in Figure~\ref{fig:ion_evo}, allowing the density peak in the blast wave to be visible at negative $y$ where it has essentially left the cloud unimpeded. Overall, the impact of the SN was very similar to that in run ION-S in the slight increase in density in the cavity walls. Very slightly different was that the low density gas left in the cavity by the supernova's wake was actually an increase in density, as it had actually been vacuum at the time of insertion.}
	\label{fig:dfb_evo}
	\end{center}
\end{figure*}

Including earlier feedback in the pre-supernova evolution had a significant effect on the overall evolution.  This was due to the actions of the stellar feedback in creating escape routes in the cloud.
Stellar feedback before the supernova was implemented as either ionisation alone or ionisation alongside stellar winds. The cluster of massive stars in the deepest regions of the cloud acted as the source for this feedback, and as such was also the location of the supernova. This meant that in both cases the supernova exploded into a cavity partially surrounded by high density molecular gas, the eroded remnants of the original cloud. In some cases the boundary from ionised gas to molecular cloud was nearly flat, giving the dense cloud a wall-like geometry. Other geometries are pillar-like structures (one prominent example can be seen in run ION in Figure~\ref{fig:ion_evo} pointing from the lower left towards the central cluster) and small droplet-like clouds completely surrounded by ionised gas (easily seen at $y \approx -20 \,\mathrm{pc}$ in DFB in Figure~\ref{fig:dfb_evo}). The state of the two simulations at the point of SN detonation can be seen in the first panels of Figures~\ref{fig:ion_evo} and \ref{fig:dfb_evo}. 

Although the central cavities appear closed, particularly in the ION simulations which appear to show two bubble-like structures, this is only a projection effect in the column density. In order to quantify the differences in the non-feedback and feedback initial conditions, we investigated what fraction of the sightlines as seen from the supernova were open, which we defined as having column densities $\le 10^{-4}$ g cm$^{-2}$. This corresponds to the minimum column density in the no-feedback (NFB-S) initial conditions as well as a critical column density allowing high-velocity outflows (see Figure~\ref{fig:absvrchange}).

In each of the three supernova simulations, the sky from the position of the progenitor sink particle was split into 768 HEALPix pixels \citep{GEA05}, the centre of each pixel defining a line of sight (LOS).  
While none of the 768 HEALPix sightlines were open in the no-feedback run (NFB-S), the previous feedback runs had 67\%  and 75\% of the sky open in the ION-S and DFB-S, initial conditions, respectively, as seen from the SN's position. 
These large portions of open sky in the feedback runs formed the preferred pathways for the SN's expansion. In ION-S, it was only low density ionised gas that was initially swept up by the ejecta. In DFB-S the situation was even more extreme -- the higher level of feedback in the original simulation led to the region around the cluster being completely vacated of gas, i.e. it fell within no gas particle's smoothing kernel.

In general the two feedback runs, ION-S and DFB-S, experienced very similar effects from the SN. The edges of the dense molecular clouds facing the SN were compressed by the shock at the point that it met them, but the large-scale and rapid destruction seen in Run~NFB-S was entirely absent. Small clouds entirely or nearly entirely overrun by the explosion were compressed inwards towards their centres. Some ablation takes place along cloud edges. The latter two effects can be seen in the long pillar reaching into the centre of the two plots in Figure~\ref{fig:ion_evo} from the bottom left. Finally, the central low density region in both runs expanded slightly. 


\subsection{Shock expansion rate}

Examining Figure~\ref{fig:rshock} shows how different the rate of the shock's expansion is depending on whether early feedback was included in the simulations. The shocks are very asymmetrical so we calculated the shock expansion rate in two directions, defined as having the highest and lowest column densities according to the HEALPix method outlined above.
The expansion rates were then calculated along the  most and least dense LOSs in order to take account of the highly inhomogeneous surroundings. The shock positions were defined by finding the position with the steepest radial entropy gradient (excluding the noisy low density cloud boundary). The bottom plot compares the column density found along each of these paths at the supernova detonation time.

The shock expanded continuously for both LOSs in the no previous feedback run, Run~NFB-S. Expansion was slow along the densest LOS and it continued to decelerate until by the final time reached in this simulation, $1.35\times 10^5\,\mathrm{yrs}$, it had only reached a distance of $6.09\,\mathrm{pc}$ from the SN progenitor's original position. Interestingly, the expansion rate along this line of sight was originally close to the Sedov-Taylor rate $r \propto t^\frac{2}{5}$ while by $10^4\,\mathrm{yrs}$ it transitioned to a slope more similar to the pressure-driven snowplough rate of $r \propto t^\frac{2}{7}$; these two slopes are respectively shown in Figure~\ref{fig:rshock} as the solid and dashed grey lines. That any match is found at all is surprising as these rates would be expected to apply only to expansion into a uniform density medium. This transition may have been driven by an increase in radiative losses as the shock moved into denser gas as seen from around $5\,\mathrm{pc}$ along this LOS in the lower plot of the figure.

The shock in Run~NFB-S moved outwards much more quickly along the least dense LOS. By the same final time it had reached a distance of $47.6\,\mathrm{pc}$ from the initial position. Our method for finding the shock position began to break down however at later times when it began to move through the outer regions of the cloud, as can be seen by the strange behaviour of the data in Figure~\ref{fig:rshock}, and as such we would not rely on it. The general behaviour of fast expansion can still be reliably taken. Along this LOS there is no clear relation to the expansion rate of either the Sedov-Taylor or the pressure-driven snowplough phases.

It is immediately apparent that the shock expansion was initially much faster in Runs~ION-S and DFB-S, where previous stellar feedback had at least partially cleared the vicinity of the supernova progenitor.  Notably in these two runs, the shock nearly stalled in the densest LOSs once a distance of around $6$ or $7\,\mathrm{pc}$ was reached, while the expansion in the least dense directions continued allowing the supernova energy (and pressure) to be released. Examining the lower plot of Figure~\ref{fig:rshock} one can see that there are very large jumps in column density at the corresponding distance, matching the point where the LOS entered one of the walls of molecular gas surrounding the central cavity. Expansion did continue beyond this time but very slowly. The least dense LOSs which passed only through ionised gas (and, as previously noted, vacuum in the central regions of DFB-S) allowed the shock to continue advancing at a rapid pace.

\section{Effect of the environment on supernova output}

Our goal is to understand how the distribution of nearby molecular gas in turn affects the distribution of the initially isotropic output from the supernova. It is clear from the previous section that the supernova expansion rate  is very different depending on the direction. Here we look to investigate how the environment affects the kinetic energy  and momentum deposition in the cloud.

\subsection{Large-scale distribution of output}

\begin{figure*}
	\begin{center}
	\includegraphics[width=0.9\textwidth]{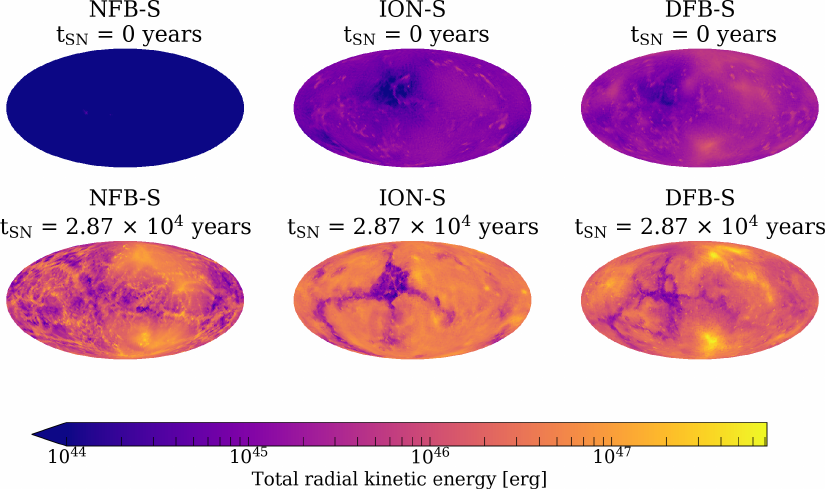}
	\caption{Distribution across the sky of radial kinetic energy in the ambient (non-supernova ejecta) gas.}
	\label{fig:ek_angular}
	\end{center}
\end{figure*}

To understand better how the supernova propagates through the cloud, 
Figure~\ref{fig:ek_angular} shows the angular distribution of kinetic energy across the sky from the position of the supernova progenitor. The radial kinetic energy--that is to say, the kinetic energy calculated using only the radial component of the velocity--was calculated for each SPH particle making up the ambient gas and then added to whichever of the 12,288 HEALPix pixels \citep{GEA05} in which it was located.

From Figure~\ref{fig:ek_angular} we can see that there was initially much more kinetic energy in the two runs which had experienced earlier feedback, as would be expected. After the SN, a large amount of kinetic energy was distributed across the sky in the no previous feedback  simulation NFB-S, in a complex filamentary structure with significant voids where little kinetic energy is deposited. Of greater importance are two large regions of high energy, and it is at these positions that the SN ejecta was able to break out of the cloud as was seen in the later times shown in Figure~\ref{fig:nfb_evo}.

The kinetic energy was distributed much more smoothly across the sky in Run ION-S where ionisation had previously cleared the inner regions. There is also a very large area of low energy located close to the centre of the plot for $t_\mathrm{SN} = 2.87\times 10^{4}\,\mathrm{years}$. Four `arms' spread from it. Small regions of low kinetic energy are scattered elsewhere across the sky. These all correspond to dense clouds of varying size and shape which have been able to shield themselves from the SN and so have not received much (or any) kinetic energy from the explosion.

Run DFB-S shows a cross-shaped structure of low kinetic energy at the later post-SN time, roughly corresponding with the large low kinetic energy region seen in ION-S. The structure is however thinner and covers less of the sky, reflecting the reality that this simulation was previously bombarded with winds as well as ionising radiation, leaving less material able to self-shield from the SN. Interestingly, there are two areas of higher kinetic energy closely corresponding those seen in NFB-S. Preferential channels for the escape of the SN ejecta and energy still exist, and since all simulations were evolved from the same initial seed it should be expected that some similarities between them will remain. 

\subsection{Local energy deposition}

\begin{figure}
	\begin{center}
		\subfloat[Run NFB-S\label{fig:absvr-nfb}]{
			\includegraphics[width=9cm]{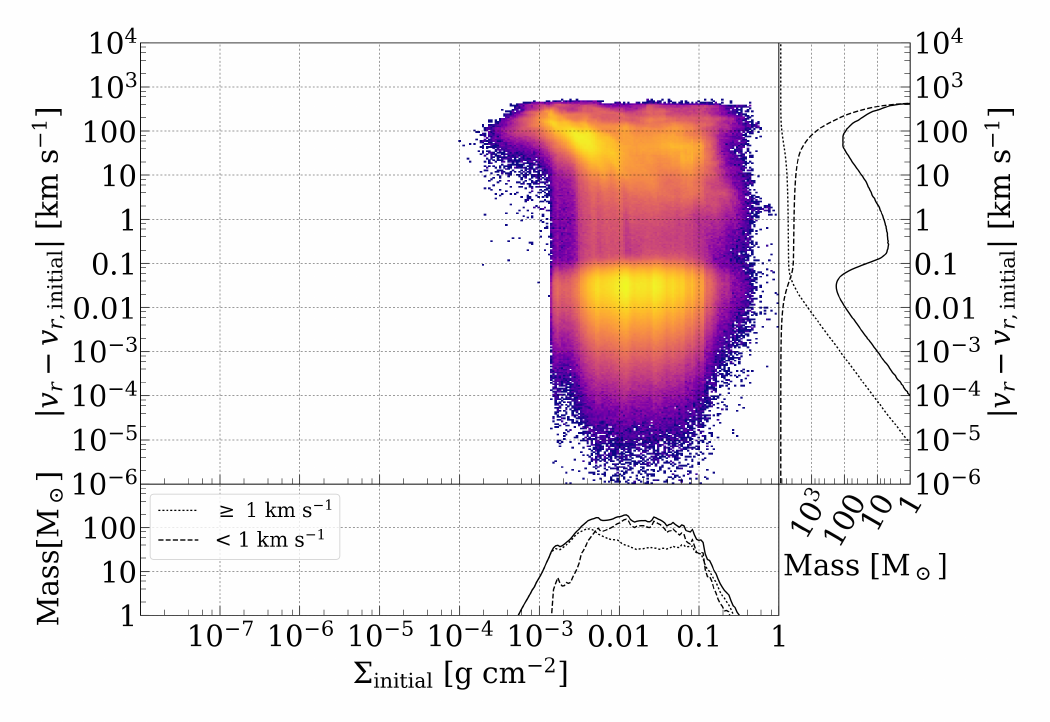}
		}\hspace{-0.5cm}
		\subfloat[Run ION-S\label{fig:absvr-ion}]{
			\includegraphics[width=9cm]{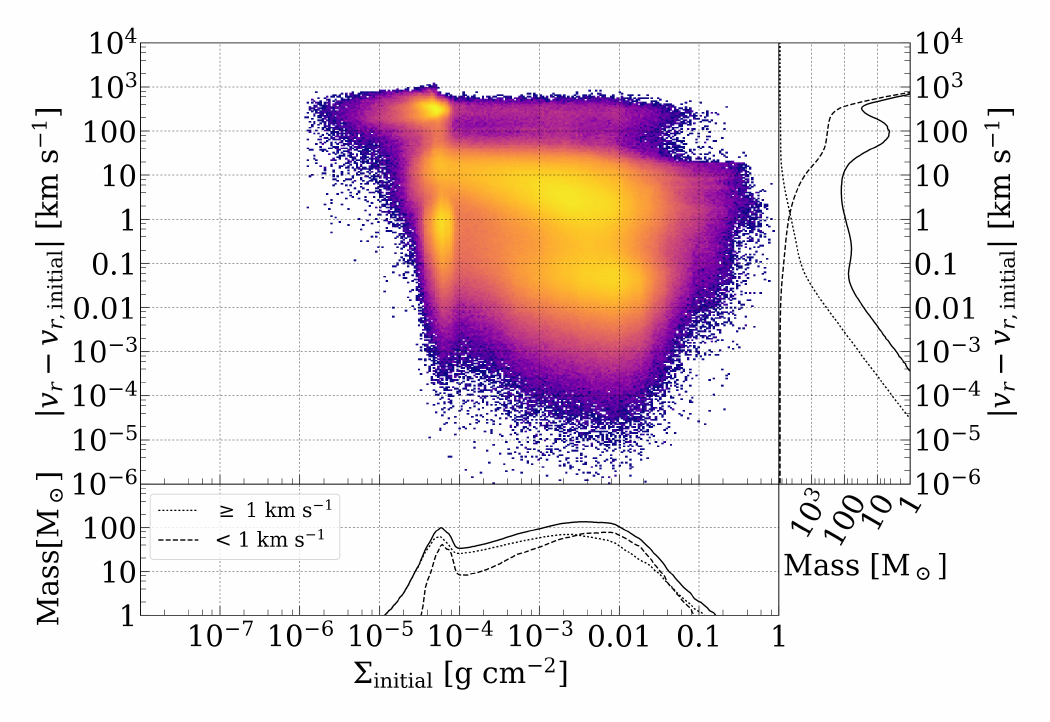}
		}\hspace{-0.5cm}
		\subfloat[Run DFB-S\label{fig:absvr-dfb}]{
			\includegraphics[width=8cm]{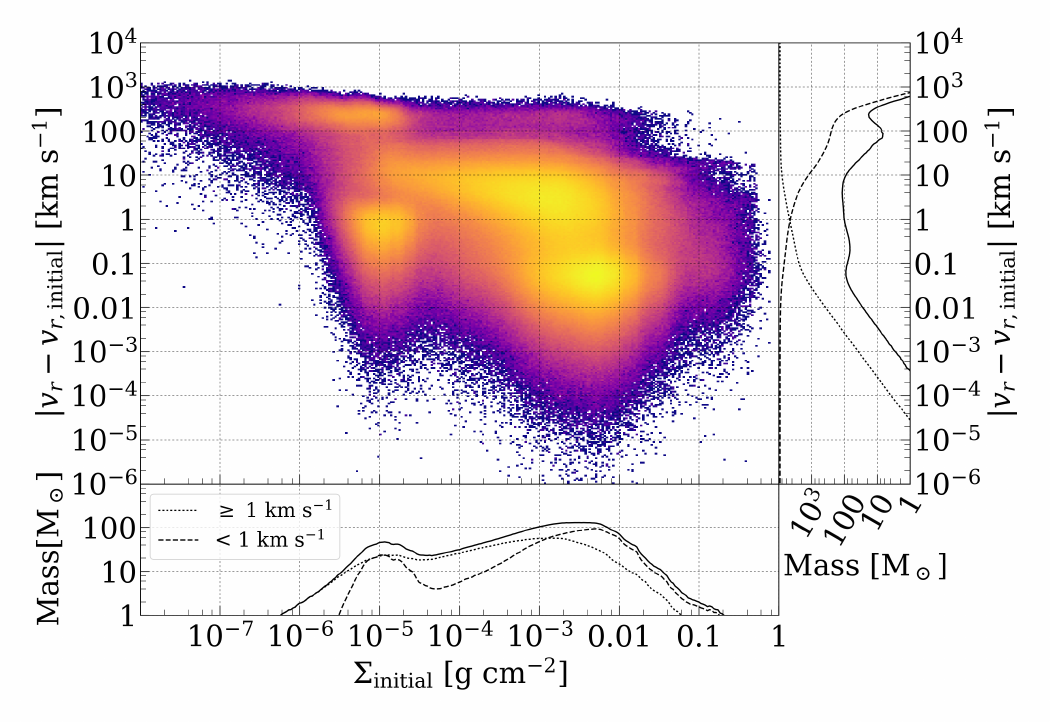}
		}
		\caption{Phase diagrams at $2.87 \times 10^4 \,\mathrm{yrs}$ showing the line-of-sight column density from the supernova and the change in radial velocity for all gas particles in the cloud (i.e. the SN ejecta is excluded). The bottom and right-hand panels contain the same information compressed to one dimension. The bottom histogram of $\Sigma_\mathrm{initial}$ shows the total and also two histograms for gas whose radial velocities changed by more or less than $1\,\mathrm{km}\,\mathrm{s}^{-1}$, itself marked on the main histogram. The right-hand histograms also show forward and reverse cumulative sums of the distribution.}
		\label{fig:absvrchange}
	\end{center}
\end{figure}

\begin{figure}
	\begin{center}
		\includegraphics[width=8.5cm]{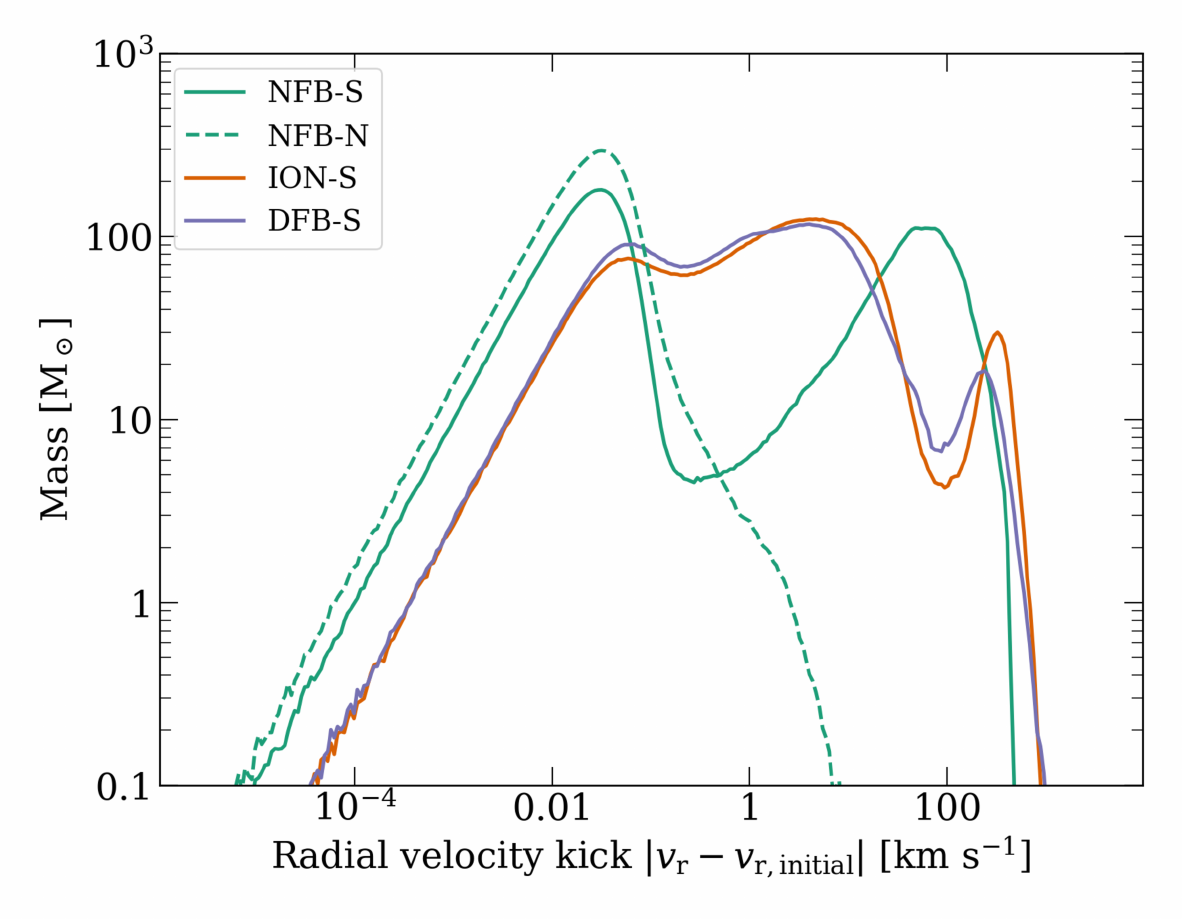}
		\caption{The distribution of mass is shown as a function of the  radial velocity kick, defined as the absolute change in a given particle's radial motion between the start of the simulation and the measurement time at $2.87 \times 10^4 \,\mathrm{yrs}$. The plot shows the comparison of the three supernova simulations and one control run without supernova explosion. The supernova generates significant kick velocities of up to several hundred km s$^{-1}$.  The high kick velocities are almost entirely from gas particles that are exposed, ie low $\Sigma \le 4 \times 10^{-3}\,\mathrm{g}\,\mathrm{cm}^{-2}$. The no previous feedback simulation has a bimodal distribution as more mass is swept up by the supernova but to lower velocities. The two cases with previous feedback has more mass given moderate (several km s$^{-1}$) but also significant peaks of mass at the highest kick velocities.}
	\label{fig:vrchange_pdf}
	\end{center}
\end{figure}

We have seen how the energy deposited in the surrounding cloud is very asymmetrical and depends on the pre-supernova structure extant in the environment. 
Globally, the three simulations' initial conditions vary by the presence and volume of escape channels carved out by earlier feedback.
In this subsection we  examine how individual mass elements were affected by the supernova and how this depended on their local shielding from the supernova.  For each gas particle we calculated the column density $\Sigma$ between the location of the particle and the SN, with this value acting as a proxy for `exposure' to the SN. Thus a low $\Sigma$ particle could be described as being exposed to the SN, independent from its distance, while a high $\Sigma$ particle was shielded from the explosion.

Figure~\ref{fig:absvrchange} shows phase diagrams for gas particles in the three simulations, plotting the `kick' $|v_\mathrm{r} - v_\mathrm{r,initial}|$ received by each particle against the initial exposure $\Sigma_\mathrm{initial}$ to the SN. The kick is simply the absolute change in the particle's radial velocity, with the SN at the origin. The initial time of the supernova explosion was at $t_\mathrm{SN} = 0$ years, while the kicks were found for $t_\mathrm{SN} = 2.87 \times 10^4$ years (the end time of DFB-S). Histograms formed by compressing the phase plot in each direction are also shown. The initial supernova particles are not shown in Figure~\ref{fig:absvrchange} as they have no values for $\Sigma_\mathrm{initial}$.

The diagram for the no previous feedback simulation, NFB-S, is the simplest to understand. The phase diagram shows two populations of particles: the particles in one received changes in radial velocity barely exceeding $0.1\,\mathrm{km}\,\mathrm{s}^{-1}$, while in the other they reached several hundred $\mathrm{km}\,\mathrm{s}^{-1}$. The high kick group was not present initially, but forms as the supernova expands and interacts with the surrounding material.  Particles across the whole range of $\Sigma_\mathrm{initial}$ were able to reach the highest velocities--as the explosion was confined within the molecular cloud; it was possible for anything nearby to become swept up in the shock irrespective of its initial environment. There is a significant increase in the particle density at lower   $\Sigma_\mathrm{initial}$ in this high-kick group, as can be seen in the corresponding histogram: the higher kick particles were more likely to be  more exposed in the initial conditions. In other words, there is still a preference for the expansion of the shock to progress along lower density paths.

The plots for the previous stellar feedback runs, ION-S and DFB-S are more complex. The two simulations show similar distributions of particles across four groupings in the phase diagrams, although the groups spread to lower $\Sigma_\mathrm{initial}$ in the dual stellar feedback run, DFB-S.  A significant fraction of the gas in both simulations is found at high initial column densities, $\Sigma_\mathrm{initial}$, and  the bulk of this gas receives only moderate kick velocities from the supernova. The particles here made up the dense remnants of the cloud between the escape channels carved out by ionization and winds. Maximum velocity kicks of this dense gas were some 10's of km s$^{-1}$ with much of it getting kicks of less than $\approx 0.1$ km s$^{-1}$.
The highest kicks extending up to nearly $10^3\,\mathrm{km}\,\mathrm{s}^{-1}$ were experienced by particles in the low $\Sigma_\mathrm{initial}$ (exposed to the SN) group. Ionised gas still untouched by the shock at $t_\mathrm{SN} = 2.87 \times 10^4$ years can be seen in the lower left group of particles, while those particles through which it has already passed can be seen directly above at the highest values for $|v_\mathrm{r} - v_\mathrm{r,initial}|$ between $100$ and $10^3\,\mathrm{km}\,\mathrm{s}^{-1}$.

The contrast between the simulation with no feedback before the SN, NFB-S, and those two with feedback, ION-S and DFB-S, is then in which material received large kicks to their radial velocities, and the magnitude of those kicks. With no  previous feedback, in NFB-S, all the gas around the supernova could receive increases to their radial velocities of up to several hundred $\mathrm{km}\,\mathrm{s}^{-1}$, though the majority of the fastest moving material was the most exposed to the SN. The reverse cumulative histogram on the right-hand panel of Figure~\ref{fig:absvr-nfb} shows that $10^3\mathrm{M}_\odot$ of the total $10^4\mathrm{M}_\odot$ in the cloud was accelerated by at least $100\,\mathrm{km}\,\mathrm{s}^{-1}$.

When ionisation and winds were allowed to shape the ISM in ION-S and DFB-S, the least exposed (highest $\Sigma_\mathrm{initial}$) gas, within dense molecular clouds, received at   most of order  $ 10\,\mathrm{km}\,\mathrm{s}^{-1}$. Only the very low $\Sigma_\mathrm{initial}$ ionised gas could be pushed to radial velocities similar to or higher than those seen in NFB-S. This can be seen clearly in  Figure \ref{fig:vrchange_pdf} that shows the mass distribution as a function of the velocity kick received from the supernova over the $ 2.87 \times 10^4$ years
of the various simulations. The no-previous feedback run shows a double peaked distribution with nearly equal parts receiving essentially no kick and kicks of some 10-100 km s$^{-1}$ with the high-kick population coming entirely from the exposed gas with low column densities to the supernova. The previous stellar feedback runs show a wider distribution of kick velocities,  
with a smaller total gas mass at  kick velocities above 1 km s$^{-1}$,  but significantly a small peak  of the mass at very high velocities of $>100$ km\ s$^{-1}$. This shows the effect of the channelling of the supernova explosion by the previous feedback such that a smaller fraction of the mass then contains a much higher fraction of the supernova's kinetic energy. This will increase the amount of energy that can escape the natal cloud in these simulations.


\subsection{Mass loss}\label{ss:mass_loss}


\begin{figure}
	\begin{center}
		\includegraphics[width=8.5cm]{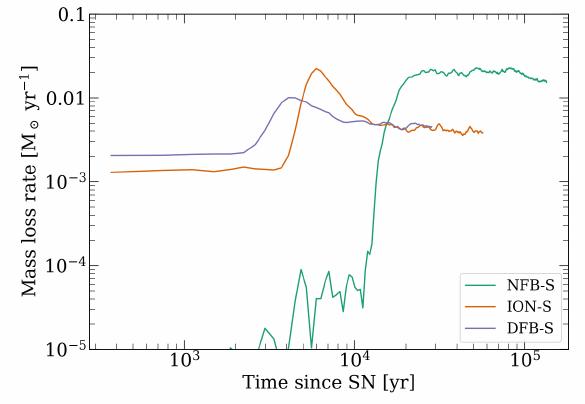}
		\caption{The  mass-loss rate,  through a sphere of 10 pc centred on the supernova progenitor, is shown for the full simulation time in each of the three runs.
		The pre-supernova feedback runs (ION-S and DFB-S) have initial mass loss rates of $\approx 10^{-3}\mathrm{M}_\odot\,\mathrm{yr}^{-1}$. 	
These  temporarily (few $\times 10^3$ years) increases to values of  $1-2 \times 10^{-2}\mathrm{M}_\odot\,\mathrm{yr}^{-1}$ as the supernova shock passes  10pc. In contrast, the  no feedback run (NFB-S) has  an essentially zero mass loss rate until the supernova shock reaches 10 pc, but then   sustains  a high mass loss rate of $\approx 2 \times 10^{-2}\mathrm{M}_\odot\,\mathrm{yr}^{-1}$ over many times $10^4$ years. This shows that the structures due to the previous feedback are able to channel the supernova outwards, allowing it to escape without significantly affecting the remaining cloud.}
		\label{fig:mass_loss}
	\end{center}
\end{figure}

The supernova's ultimate impact on the larger scale environment depends on how much mass and  energy escapes the natal molecular cloud.  We measure the mass loss as the time derivative of the mass contained within 10 pc of the supernova, as seen in Figure~\ref{fig:mass_loss}.  All three simulations show initially low mass loss rates with the non-previous stellar feedback run (NFB-S) having effectively zero mass loss rates whereas the ionisation (ION-S) and dual feedback runs (DFB-S)  
have initial mass loss rates of  $1-2 \times 10^{-3} \mathrm{M}_\odot\,\mathrm{yr}^{-1}$. 

The mass- loss rates increase significantly as the supernova shock passes reaches $10$ pc (see Figure~\ref{fig:rshock}). This occurs after 2600 years in the DFB-S simulation with a peak mass loss rate of $1.0 \times 10^{-2}\mathrm{M}_\odot\,\mathrm{yr}^{-1}$ at 4100 years. In the ionisation simulation (ION-S) the supernova shock required longer (3700 years) to reach 10 pc but then
produced a higher peak mass loss rate of $2.2 \times 10^{-2}\mathrm{M}_\odot\,\mathrm{yr}^{-1}$ at 6000 years. Finally, in the no previous feedback run (NFB-S), the supernova shock does not reach 10 pc until 11500 years but then produces a high, and sustained mass loss rate of $2 \times 10^{-2}\mathrm{M}_\odot\,\mathrm{yr}^{-1}$ over $10^5$ years. 

It is clear that the existence of significant holes in the cloud due to the previous feedback events are able to channel the supernova shock more quickly out of the cloud, and hence less mass is lost from the inner 10 pc than in the case where the supernova shock has to propagate through a pristine molecular cloud. Even though more mass is lost from the NFB-S case,  it should be noted that this cloud has not already lost significant mass from any previous feedback event. The higher mass loss from the NFB-S simulation ($\approx 2 \times 10^3 \mathrm{M}_\odot$) succeeds in reducing the cloud mass to be comparable  with the initial mass in the ION-S cloud. Initial masses for the three clouds were NFB-S: $4604 \mathrm{M}_\odot$; ION-S: $2620 \mathrm{M}_\odot$; DFB-S: $1290 \mathrm{M}_\odot$. 

We can conclude that clouds with previous stellar feedback are more porous, with pre-existing channels allowing the supernova explosion to escape more quickly, and without sweeping up as much mass than would be the case when no previous stellar feedback occurred. This is important in allowing the supernova to escape the natal environment and
this affect the larger scale in the galactic interstellar medium.

\subsection{Energy loss}


\begin{figure}
	\begin{center}
		\includegraphics[width=0.48\textwidth]{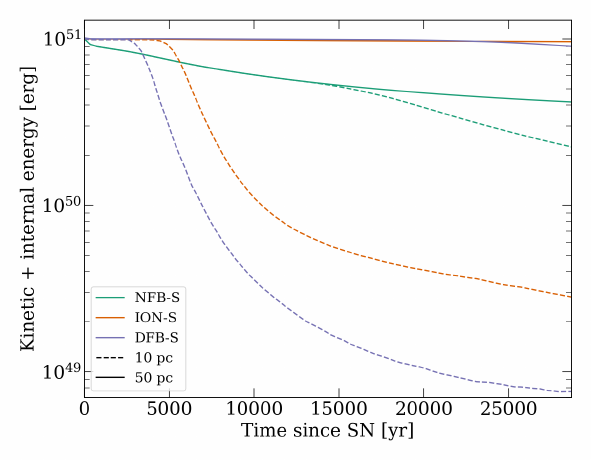}
		\caption{This figure plots against time the summed kinetic and internal energy contained within spheres centred on the supernova of radius 10 and 50 pc. Energy on small scales (10 pc) is lost most quickly in the feedback runs, ION-S and DFB-S, as the ejecta and swept up material carried energy outwards to larger distances. However, on larger scales (50 pc) more was lost in the feedback run in which the energy was thermalized in the shock's interaction with dense cloud material and then lost radiatively.}
		\label{fig:energy_loss}
	\end{center}
\end{figure}

\begin{figure*}
	\begin{center}
	\includegraphics[width=\textwidth]{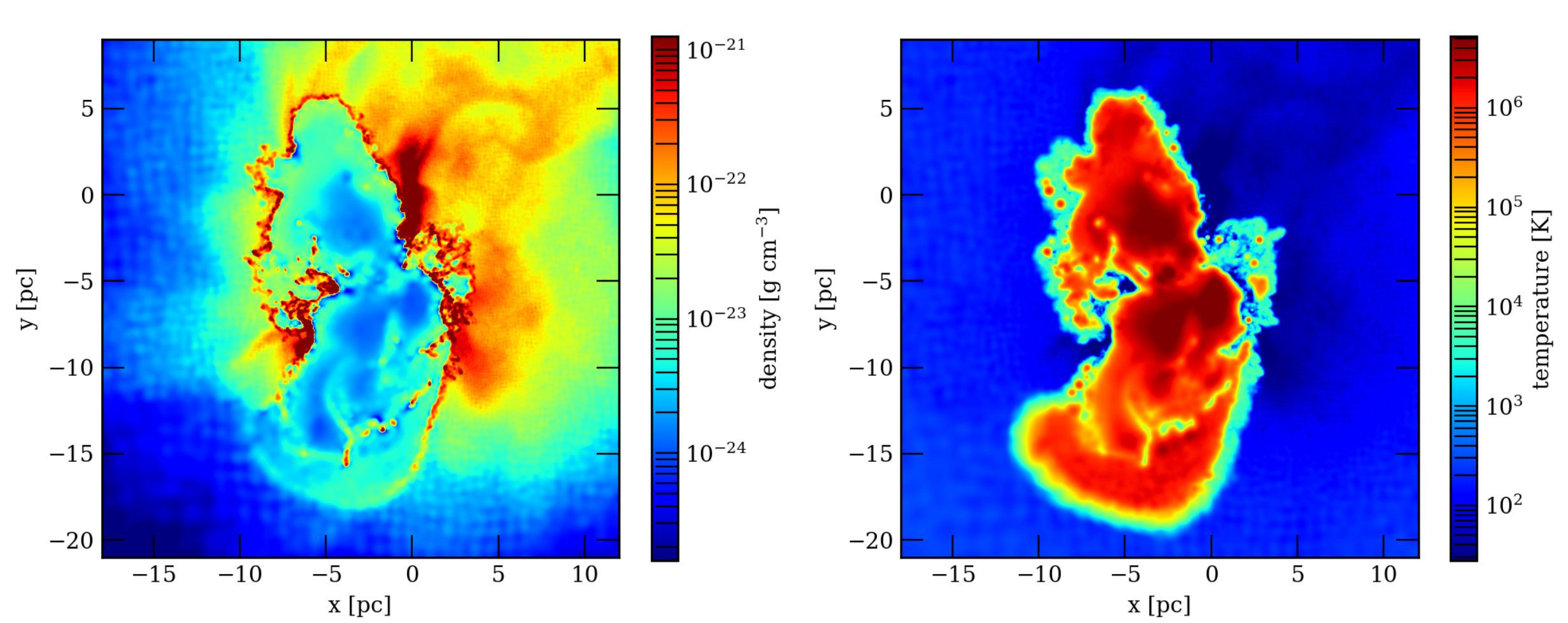}\hspace{-0.5cm}
	\includegraphics[width=\textwidth]{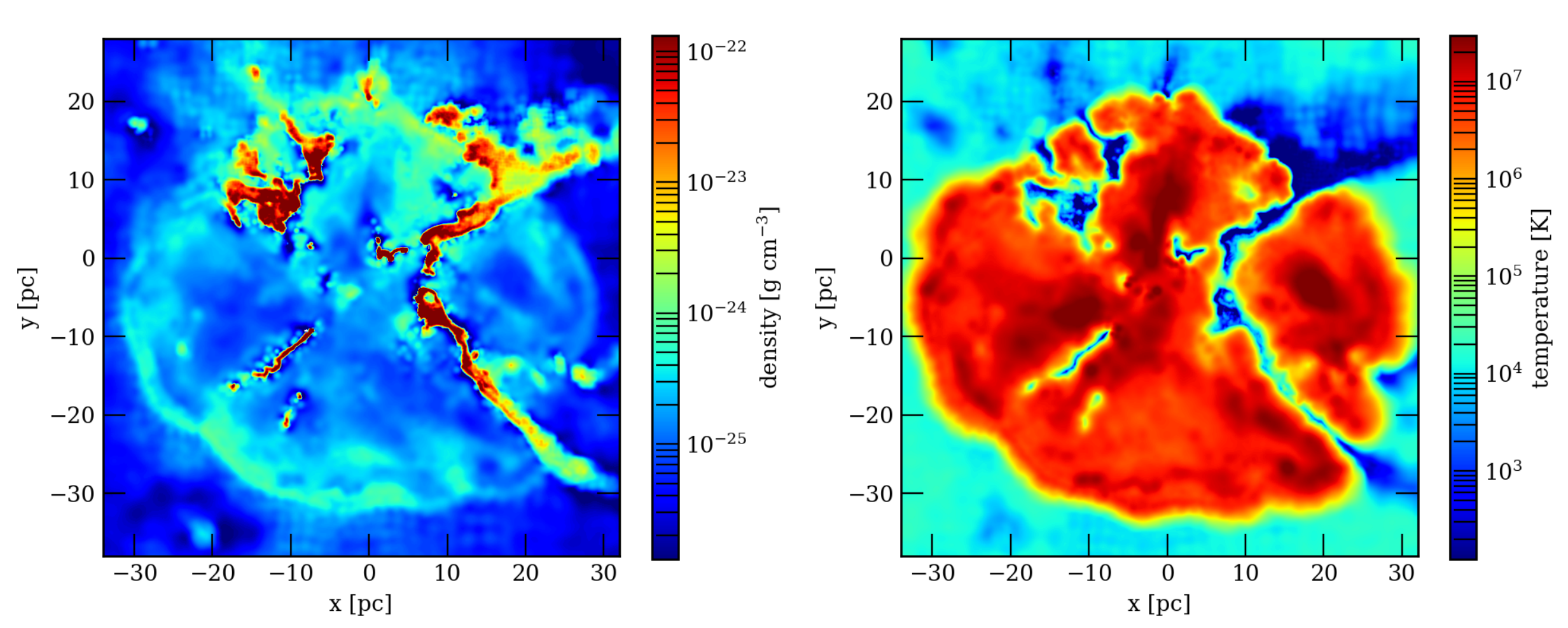}
	\caption{A cross section of densities (left) and temperatures (right) of run NFB-S (top) and ION-S (bottom) are shown at $2.87 \times 10^4$ years after supernova explosion. The high-density gas is from the pre-supernova environment and remains cool whereas the cavities are  filled up of low density gas at temperatures in excess of $10^7$ K. Over-cooling of the supernova remnant is not occurring in this simulation. The slightly lower temperatures in the no feedback simulation (NFB) are due to the  larger amount of mass that has been swept up  in the explosion. The limit of the SN explosion is clear in both cases as is the effect of the ionising feedback in creating a large cavity for the SN to explode into. }
	\label{fig:SNtemps}
	\end{center}
\end{figure*}

As well as examining how mass was removed across the three simulations, it is also beneficial to look at how energy was lost. The sum of the kinetic and internal energies is plotted as a function of time for the three supernova simulations (Figure~\ref{fig:energy_loss}). The energy is calculated within radii of 10 and 50 pc, the latter intended to capture the supernova even after it has escaped the cloud.  The initial energy deposition in the supernova was set at $10^{51}$ ergs. 

The combined kinetic and internal energies within 10 pc are all seen to decrease once the supernova shock passes this distance, as seen in Figure~\ref{fig:rshock} and 
Figure~\ref{fig:mass_loss}.  This occurs first for the dual feedback run (DFB-S), then later for the ionisation feedback run (ION-S) and last for the no previous feedback run (NFB-S).  Once the shock has passed through  the inner regions of the cloud, the energy drops rapidly leaving the remnant, well shielded portions  of the cloud largely unaffected. This is aided by the existence of the channels eroded into the cloud due to the previous feedback.

In contrast, the supernova shocks do not reach 50 pc over the timescale of the simulations, so the combined kinetic and internal energy then measures the energy conservation of the supernova as it expands in the cloud. What is most important, is that the combined kinetic and internal energies are near constant for the two previous feedback simulations (DFB-S and ION-S), whereas there is a significant decrease for the no previous feedback simulation (NFB-S). This occurs from the initial stages of the simulation, well before the shock has reached 10pc. Without any previous feedback, there are no well-formed channels in the cloud through which the supernova can escape. Instead it needs to create these channels in the weakest regions of the cloud with low column densities. This creating of the outflow channels involves  sweeping up more of the gas and distributing the supernova energy into a larger mass distribution. When this occurs, more of the internal energy is able to escape through radiation from this larger, and hence cooler, mass distribution. Hence in Figure~\ref{fig:energy_loss} the no previous feedback simulation has already lost over half its combined kinetic and internal energies in the first 25000 years from the explosion. 
The no previous feedback simulation was followed much further than the other simulations. Over the full 120000 years, we see that the internal energy decreases to $\approx 5$ per cent of its peak value whereas the two feedback runs show negligible loss of internal energy over their full runs. 

This issue is generally referred to as over-cooling and is seen as a resolution limitation in cosmological simulations. Here we can see that even at high resolutions this can be an effect but that with realistic initial conditions where previous stellar feedback has created outflow channels in the cloud, the supernova can readily escape the dense regions without sweeping up too much mass and hence suffering from over-cooling and an excessive loss of energy. 

 It is worth noting that the supernova remnant remains very hot even in the cases where significant energy is lost. Figure~\ref{fig:SNtemps} shows a cross section of the density and temperature in the region for the no feedback (NFB-S) and the ionisation (ION-S) cases. Temperatures in excess of $10^6$ to $10^7$ K are present in the feedback bubbles. The remaining dense pre-supernova gas is very cold and the contrast between the two highlights how this structure co-exists in the region while the supernova eject escapes through the weak points in the cavity walls. The temperatures are slightly lower in the feedback bubble  for the no previous feedback case as the SN had to sweep up more material in its path and hence increase the cooling. The ionisation and dual feedback simulations have the highest temperatures in the feedback bubble due to the lower densities in the pre-supernova cavity created by the earlier feedback mechanisms.

\section{Discussion}

One of the surprising results from these simulations is that the supernova explosion, with many orders of magnitude more energy than the full gas cloud with its embedded stellar cluster, does not completely destroy the cloud. Instead, it creates channels, or employs already created channels due to previous stellar feedback, and largely escapes the dense cloud. The fact that the cloud can channel the supernova outflow is due to the large gas pressure in the cloud such that the dense filaments are able to withstand the onslaught and shield much of the cloud. In the case where channels have already been formed by previous feedback, the supernova can quickly escape the inner regions which reduces the thermal and kinetic pressures. It is clear from the kick velocities received from the supernova that the dense shielded gas is largely unaffected by the supernova. Feedback hence has little effect on nearby ongoing star formation.

In contrast, these simulations show that the exposed gas at low column densities is very much susceptible to the effects of the feedback. Exposed gas within the clouds at column densities $\Sigma < 10^{-4}$~g~cm$^{-3}$ can be swept up and removed from the cloud. With pre-existing channels created by stellar feedback, the mass swept up is reduced which allows more of the supernova energy to escape unaltered by the dense cloud. The radiative losses from cooling by the denser gas is reduced and the full impact of the supernova is permitted to leave the cloud and directly impact on the large scale environment of the galaxy. It can thus have a significant effect on star formation on larger scales by energising of the interstellar medium of the galaxy.

 It should be noted that the results presented here are for moderate mass clouds of only $10^4 {M}_\odot$  and that larger clouds with higher escape velocities are more robust against earlier forms of feedback \citep{DEB2012}. Such clouds may have more contained inner cavities from previous feedback events and hence act to better constrain the supernova explosion to act within the cloud, and hence comparable to the no feedback run presented here. The inner regions where ionisation and winds will have created cavities, will most likely be comparable to the previous feedback runs in our $10^4{M}_\odot$  cloud.

Of further note is the effect of resolution in simulating feedback from supernova into larger scale simulations. Generally such simulations will  be resolution limited and unable to resolve the structure in the pre-supernova cloud, be it generated by turbulence or earlier feedback events. The surroundings will then be typically more uniform and of lower median densities allowing the supernova to sweep up more of the material. This would then result in a stronger local effect of the supernova feedback, destroying the local cloud, but higher cooling rates and energy losses and typically a lower effective feedback on larger scales. This was evident in our earlier simulations where we did not include particle splitting and should remain a concern to all studies that include feedback into the interstellar medium.

\section{Conclusions}

We have shown that resolving the natal environment, and including previous stellar feedback, is essential to accurately model the impact of supernovae on their immediate and even larger scale environment. Realistic natal environments include significant dense gas structures and filaments which can act to channel the
supernova outflow through the weaker regions of the cloud with lower column  densities. The supernova's energy is very preferentially deposited in these regions with lower column densities, ie less shielded from the supernova.
This asymmetrical ejecta leaves the dense, shielded, regions largely unaffected by the supernova, as has been seen in the case where only stellar feedback is modeled \citep{DBCB2005,Dale2017}.

When previous stellar feedback events are included, the pre-existing channels are well-formed, allowing a more rapid and efficient escape of the  supernova energy and ejecta, with less gas mass being swept into the shock.  The supernova energy is then deposited in less mass, allowing the supernova shock to leave the cloud faster, and to lose less energy due to cooling processes that would occur in a higher mass ejecta. We have only considered relatively moderate mass clouds ($10^4 M_{\odot}$) that are most susceptible to pre-supernova feedback \citep{DEB2012}. Supernova in larger mass clouds is likely to have central regions similar to our previous feedback runs while the outer areas of the cloud would more closely resemble the no-previous feedback case. 

These results imply that modelling the natal environment of supernovae is crucial in order to model their energetic coupling with the ISM at the correct scales.. When stellar feedback
is included, most of the energy of the supernova can escape its  natal molecular cloud and can thus impact lower density gas at larger distances in the interstellar medium. This will have important consequences for galaxy formation and evolution.

\section*{Acknowledgements}

WEL and IAB acknowledges funding from the European Research Council for the FP7 ERC advanced grant project ECOGAL. This work used the compute resources of the St Andrews HPC Cluster Kennedy, and the DiRAC Complexity system, operated by the University of Leicester IT Services, which forms part of the STFC DiRAC HPC Facility (www.dirac.ac.uk). This DiRAC equipment is funded by BIS National E-Infrastructure capital grant ST/K000373/1 and  STFC DiRAC Operations grant ST/K0003259/1. DiRAC is part of the National E-Infrastructure.

\end{document}